# Title: Ultrafast Generation and Control of an Electron Vortex Beam via Chiral Plasmonic Near Fields

**Authors:** G. M. Vanacore[1,*,†], G. Berruto[1,*], I. Madan[1], E. Pomarico[1], P. Biagioni[2], R. J. Lamb[3], D. McGrouther[3], O. Reinhardt[4], I. Kaminer[4], B. Barwick[5], H. Larocque[6], V. Grillo[7], E. Karimi[6], F. J. García de Abajo[8,9], and F. Carbone[1]

**Affiliations:**

[1]Institute of Physics, Laboratory for Ultrafast Microscopy and Electron Scattering (LUMES), École Polytechnique Fédérale de Lausanne, Station 6, CH-1015 Lausanne, Switzerland

[2]Dipartimento di Fisica, Politecnico di Milano, Piazza Leonardo da Vinci 32, 20133 Milano, Italy

[3]SUPA, School of Physics and Astronomy, University of Glasgow, Glasgow G12 8QQ, UK

[4]Faculty of Electrical Engineering and Solid State Institute, Technion, Haifa 32000, Israel

[5]Ripon College, 300 W. Seward St., Ripon, WI 54971, United States

[6]Department of Physics, University of Ottawa, 25 Templeton St., Ottawa, Ontario, K1N 6N5 Canada

[7]CNR-Istituto Nanoscienze, Centro S3, Via G Campi 213/a, I-41125 Modena, Italy

[8]ICFO-Institut de Ciencies Fotoniques, The Barcelona Institute of Science and Technology, 08860 Castelldefels (Barcelona), Spain

[9]ICREA-Institucío Catalana de Recerca i Estudis Avançats, Passeig Lluís Companys 23, 08010 Barcelona, Spain

*Authors contributed equally.

†Correspondence to: giovanni.vanacore@epfl.ch

**Introductory paragraph**: Vortex-carrying matter waves, such as chiral electron beams, are of significant interest in both applied and fundamental science. Continuous wave electron vortex beams are commonly prepared via passive phase masks imprinting a transverse phase modulation on the electron's wave function. Here, we show that femtosecond chiral plasmonic near fields enable the generation and dynamic control on the ultrafast timescale of an electron vortex beam. The vortex structure of the resulting electron wavepacket is probed in both real and reciprocal space using ultrafast transmission electron microscopy. This method offers a high degree of scalability to small length scales and a highly efficient manipulation of the electron vorticity with attosecond precision. Besides the direct implications in the investigation of nanoscale ultrafast processes in which chirality plays a major role, we further discuss the perspectives of using this technique to shape the wave function of charged composite particles, such as protons, and how it can be used to probe their internal structure.

**Main Text:** The quantum wave nature of both light and matter has enabled several tools to shape them into new wave structures defined by exotic non-trivial spatio-temporal properties (1). Among these techniques, the impartment of a vortex onto their transverse phase profile is showing a significant impact in both applied and fundamental science. Vortices are stagnant points at which the phase is undefined, while along a contour around them it varies by an integer multiple $m$ of $2\pi$. The magnitude and sign of the quantization number $m$, also known as the topological charge, indicate the amount and the handedness, respectively, of the phase cycles surrounding the vortex (2). If the singularity is located along the wave's central axis $z$, it is simply accounted for by a phase factor $\exp(im\phi)$ in the wave function, where $\phi$ is the azimuthal angle. Provided that the modulus of the wave is circularly symmetric, then the presence of this term causes the wave to carry orbital angular momentum (OAM), as it effectively becomes an eigenstate of the z component of the OAM operator with eigenvalue $m\hbar$, where $\hbar$ is the reduced Planck constant. Vortices have been extensively studied in optical waves within the context of classical and quantum communications (3-5), optical trapping (6,7), quantum entanglement (8,9), and nanostructured plasmonic devices (10). Vortices in electron waves have also become increasingly present in modern science, especially in electron microscopy (11-14). Combining the OAM of the vortex beam with the charge of the electrons results in an increased sensitivity to local magnetic properties (15,16). Both optical and electron beams that carry OAM can be produced by means of passive devices that directly modify their wave structure, such as spiral phase plates (17), along with amplitude (18) and phase holograms (19,20). The electron's charge also allows obtaining a vortex by other methods, such as magnetic needles (21), programmable electrostatic displays (22), and tunable electrostatic phase devices (23). Alternative approaches relying on the discrete exchange of OAM between interacting relativistic electrons and optical electromagnetic waves have also been theoretically proposed (24,25). However, all the experimental implementations have been so far limited only to continuous wave vortex electron beams, restricting the range of possible applications mainly to the investigation of ground-state processes, whereas a huge amount of information resides in the exploration of the non-equilibrium dynamics.

In this *Letter*, we experimentally show that ultrafast vorticity can be imparted to charged particles by the interaction with strong local electromagnetic fields sustained by collective electronic modes. This is achieved by having ultrashort single-electron wavepackets interacting with an optically-excited femtosecond chiral plasmonic near field. This interaction induces an azimuthally varying phase shift on the electron's wave function, as mapped *via* ultrafast transmission electron microscopy (26,27). With respect to static approaches using passive phase masks, this method offers a higher degree of scalability to small length scales and a highly efficient dynamic phase control, as inherited from the ability to manipulate the ultrafast plasmonic field. We demonstrate such level of phase manipulation using a sequence of two phase-locked light pulses to coherently control the properties of the vortex beam with attosecond precision. Our experimental results are described by means of a general theoretical framework that can be applied also to other charged particles, and thus potentially to other matter waves, such as proton beams. The ability to add OAM onto the latter could have fundamental implications on open questions in hadronic physics, where OAM-carrying protons could be used to address the origin of their spin (proton spin puzzle) (28).

Several recent studies have shown that a strong longitudinal phase modulation of the electron wave function can be achieved by the interaction with intense optical fields (29-31). Given that OAM impartment relies on the modulation of an electron's transverse phase profile, a crucial aspect for the experimental realization is the ability to synthesize an optical field with a well-defined chirality that would extend over the transverse plane and be efficiently confined on the length scale defined by the electron coherence (on the order of ~ 1 µm for fs- electron pulses). The usually adopted configurations are however not suitable for this task since the spin of the photon, as defined by its circular polarization, cannot be directly mapped into the OAM of the electron in a simple electron-photon interaction process, such as in inverse transition radiation. To fulfill these requirements the interaction needs to rely on the interruption of light propagation by a scattering object, or for example be mediated by a spatially-confined OAM-carrying optical field. This condition can be generated by the excitation of chiral surface-plasmon polaritons (SPPs) (32-34), which relies on the spin-to-OAM conversion from circularly-polarized light in non-paraxial scattering and is a clear manifestation of the strong spin-orbit coupling of light when confined to subwavelength scales (33,34). When a circularly-polarized light beam encounters a properly designed nanoscale cavity, it is able to scatter into in-plane SPPs with a nonzero topological charge and helical phase distribution (33). For a cylindrically-symmetric structure, the conservation of the total angular momentum quantum number restricts the plasmon OAM to be only $+\hbar$ or $-\hbar$ according to the helicity of the incident light plane-wave. Arbitrarily large values of plasmon OAM are instead possible by using incident light with high vorticity (e.g., Bessel beams).

We generate a chiral plasmonic field by illuminating a nanofabricated hole in a Ag-film deposited on a $Si_3N_4$ membrane (diameter ~ 0.8 µm) with circularly or partially-circularly polarized light carrying energy of $\hbar\omega = 1.57$ eV per photon (see Fig. 1**a**). Surface-plasmon polaritons are generated at the hole edge and radially propagate away within the $Ag/Si_3N_4$ interface with a phase distribution that can be directly mapped by means of a nonlocal holographic method (35). The latter directly draws from a recently developed approach used to homogeneously modulate the phase of the electron wave function by means of a semi-infinite light field obtained when a nearly-plane-wave light beam is reflected from a continuous Ag layer (opaque to the light but semi-transparent to the electrons) (31). In the presence of a nanostructure, this effect would generally coexist with electron scattering from the local field of the photo-excited SPPs propagated at the $Ag/Si_3N_4$ interface, which are nonetheless phase locked with the light field reflected at the Ag surface. Despite the two fields being localized at different positions along the *z* axis, the electron wave-packet, while propagating through the structure, can interact with both of them at a temporal separation which is smaller than its longitudinal temporal coherence. The resulting position-dependent interference between the propagating light field and the SPP field gives rise to a spatially oscillating field amplitude that directly translates into a modulation of the electron wave function that can be imaged in real space (see Fig. 1**b**). This nonlocal interference thus allows us to directly access the phase distribution of the plasmon field itself. This is realized by energy-filtered imaging, where we selectively collect only those electrons that have exchanged energy with the optical fields.

The resulting measured image for an elliptically polarized optical illumination when electron and light wavepackets are in temporal coincidence, is shown in Fig. 1**c** and displays the characteristic spiral phase pattern of a chiral SPP. This pattern is in agreement with the finite-difference-time-domain (FDTD) simulations shown in Fig. 1**e** (further details are given in section S3 of the SI). The simulated phase profile of the electric field along the *z*-axis (Fig. 1**e**) exhibits a clear spiral distribution that resembles the experimental image. Moreover, the simulations clearly show that the SPP field displays in the center of the hole a phase singularity of topological charge $m = 1$. This suggests that a $2\pi$ phase shift can be imprinted onto the electron's wave function, provided that it coherently encloses the hole upon propagating through the device.

Further insights can be obtained through a semi-analytical theory. Upon propagation through the plasmon-supporting structure, the incident electron wave function $\psi_{inc}(R, \phi, z, t)$ gains inelastically scattered components:

$$\psi_\ell(R,\phi,z,t) = \psi_{\text{inc}}(R,\phi,z,t) J_\ell(2|\beta|)\exp\bigl(i\ell\arg\{-\beta\} + i\ell\omega(z/v - t)\bigr), \qquad (1)$$

labeled by the integer $\ell$, which corresponds to the number of plasmons absorbed/emitted by a single electron. A detailed, self-contained derivation of this expression is given in Ref. (31). Here, $R$ and $\phi$ are the transverse cylindrical coordinates centered at the hole, $v$ is the electron velocity, and $J_\ell$ is the $\ell^{\text{th}}$-order Bessel function of the first kind. All of these scattered components are fully described in terms of a single complex function $\beta(R,\phi)$, which captures the interaction of the electron with the light and SPP fields. As detailed in section S2 of the SI, the interaction strength is determined by the electric field component along the $z$ axis (31), which in our case can be written as $\beta \sim A + B\exp(ik_{SPP}R)$, where the first term represents the interaction of the electrons with the semi-infinite light field (reference), while the second term defines the interaction with the SPP field propagating with a wave vector $k_{SPP}$ (signal). Direct comparison between theory and experiment can be obtained for the real-space intensity distribution $I_{\text{inelastic}}(R,\phi) = \sum_{|\ell|=1}^{+\infty}|\psi_\ell|^2$ shown in Fig. 1**d**, which nicely reproduces the experimental data.

This one-to-one correspondence between the complex optical field of the SPP and the transverse distribution of the electron wave function is further evidenced in Fig. 2 by comparing experiments and FDTD simulations for both circularly-polarized and linearly-polarized light illumination. In Fig. 2**a** and 2**f** we show the energy-filtered images of the inelastically scattered electron distribution inside the hole for both polarization states. These directly correlate with the simulated spatial distribution of the modulus of $\beta$ reported in Fig. 2**b** and 2**g**. A zero-field region with cylindrical or mirror symmetry is visible inside the hole when circularly- or linearly-polarized light is used, respectively. The spatially oscillating pattern measured outside the hole for the two polarizations (Fig. 2**c** and 2**h**) is instead a direct representation of the spatial distribution of the phase of $\beta$ within the $xy$ plane (see simulations in Fig. 2**d** and 2**i**). Because the plasmon order $\ell$ also strongly defines the degree of interaction with the SPP field, the spatial distribution of the scattered electrons can be significantly modulated by $\ell$. This can be clearly seen in the *space-energy* maps shown in Fig. 2**e** and 2**j** for circular and linear polarizations, respectively, where the zero-field region spreads in size with an increasing number of absorbed/emitted SPPs.

We note that, for incident light with circular or partially-circular polarization, the phase of the interaction strength $\beta$ evolves nearly linearly with $\phi$ (i.e., $\arg\{\beta\} \propto \phi$, see Fig. S3 in the SI). This linear dependence translates into an overall phase factor $\exp(\pm i\ell\phi)$ in the wave function $\psi_\ell$, which therefore carries an OAM of $\pm\ell\hbar$ (see also schematics in Fig. 3**a**). On a fundamental level, the transfer of the above phase profile relies on the impinging light field creating plasmons with OAM of $\pm m\hbar$. In our case $m$ is equal to 1 due to the axial symmetry of the circular hole and to the $\pm\hbar$ units of the spin angular momentum carried by the circularly polarized Gaussian beam illuminating it. In a general case, the exchange of $\ell$ plasmons with the electron would thereafter result in a net transfer of $\pm\ell m\hbar$ units of OAM. The use of non-axially-symmetric structures sustaining chiral SPPs with arbitrarily large values of $m$, such as spiral plasmonic lenses (33,36,37), is therefore a promising route to create high-OAM vortex electron beams. Because of the linearity of the interaction, our approach works at any value of the incident light intensity, and thus it is intrinsically robust with respect to the external parameters. Naturally, the efficiency of the process (i.e., the maximum $\ell$ that can be efficiently reached) might be different according to the extent of the photoexcitation.

To further verify the OAM transfer, we also monitored the far-field images of the electron beam in the transversal-momentum space $(k_x, k_y)$, and determined the changes produced by the presence of illumination. This is shown in Fig. 3**c** and 3**i** for a circularly polarized light beam in combination with a sample orientation that vanishes the electron-light interaction while maintaining a significant electron-plasmon coupling (see section S2 of the SI). Comparison between the reciprocal-space map measured under these conditions (Fig. 3**c** and 3**i**) and that acquired before optical illumination (Fig. 3**b** and 3**e**) reveals a characteristic destructive interference region at the electron beam center, which attests the presence of a phase singularity in its transverse phase profile. We attribute the origin of this behavior to the chiral near field created at the hole, which dominates the coupling with the electron wavepacket (see Fig. S6 in the SI) and imprints on it a $2\pi\ell$ phase shift as inherited by the circularly-polarized optical illumination. To confirm that the reciprocal-space distribution of the electron beam is indeed determined by the phase profile of the optical field at the hole, we have performed additional experiments using a linearly-polarized light beam. This is shown in Fig. 3**d** and 3**j**, where a characteristic 2-lobed electron beam (38) is formed as a result of the uniform $\pi$ phase shift generated across the hole (see Fig. 2**i**).

Within this scenario, the momentum-resolved electron wave function $\Psi_\ell(k_x, k_y)$ can be calculated as the Fourier transform of the real-space wave function $\psi_\ell(R, \phi)$:

$$\Psi_\ell(k_x, k_y) = \int_{-\pi}^{\pi} d\phi \int_{0}^{+\infty} RdR \exp(-i\mathbf{k}_\perp \cdot \mathbf{R})\psi_\ell(R, \phi) \qquad (2)$$

where $\mathbf{k}_\perp = (k_x, k_y)$. Here, $\psi_\ell(R, \phi)$ is given by Eq. (1), where the incident wave function $\psi_{\text{inc}}(R, \phi)$ has a lateral extension determined by the electron transverse coherence (~ 0.7-1 μm), and the interaction strength $\beta$ originated at the hole is extracted from FDTD simulations where the actual spatial distribution of the near field is considered (see Fig. S6 in the SI). The total Fourier-plane intensity is then obtained as: $I_F(k_x, k_y) = \sum_{\ell=-\infty}^{+\infty} |\Psi_\ell(k_x, k_y)|^2$. The calculated maps, where a small momentum broadening is also included to take into account the experimental resolution, are shown in Fig. 3**e-g** and correctly reproduce the doughnut-shaped and the 2-lobed probability distribution observed in the experiment as induced by the chiral and linear transverse phase modulation, respectively. Our experimental and theoretical results are also in excellent agreement with standard electron vortex beam calculations (see details in section S4 of the SI). In the latter case, the effect of the chiral near field on the electron wave function is simulated by the interaction of a plane wave with a spatially-confined spiral phase plate of order $\ell$, whose extension is determined by the hole diameter (see Fig. S8 in the SI).

An extremely interesting feature of our approach is the ability to dynamically control the vortex beam properties. This is possible thanks to the high degree of tunability of the optical field, which allows to coherently modulate the longitudinal and transverse phase profile of the electron wave function. This is achieved by manipulating the phase distribution of the plasmonic near field with attosecond precision using a sequence of two phase-locked 55-fs light pulses with a relative delay, $\Delta t$, between them varied in steps of 334 as (see also Ref. (31)). As schematically depicted in Fig. 4**a**, this three-pulse scheme allows us to finely control the relative phase of the two optically-excited plasmons. In the first implemented configuration, we have used two elliptically polarized light pulses with parallel major axes. The initial delay was set to $\Delta t^* = 85$ fs, longer than their cross-correlation, in order to minimize the optical interference between the two pulses, which remained nevertheless below a 5 x $10^{-2}$ level. When varying the delay time between the two optical pulses, the observed time-dependent evolution of the plasmonic pattern is a sequence of constructive and destructive interference between the optically-generated plasmons propagating at the Ag/Si$_3$N$_4$ interface (see Fig. 4**b-d**). The result is a coherent modulation of the intensity and of the spatial periodicity of the plasmonic fringes with a temporal period given by the optical cycle of ~ 2.67 fs (see Fig. 4**e-f**, Supplementary Movie S1, and calculations in Fig. S4 in the SI). This effect, which cannot be related to a simple optical interference within the incident light field, can be understood as the coherent interaction between two spatially- and temporally-localized plasmon wavepackets. In a second configuration, the elliptical polarization of the two light pulses is arranged such that their major axes are perpendicular to each other. In this way, we effectively have a dynamic handle on the helicity of the phase singularity created at the center of the nanofabricated hole. This is shown in Fig. 4**g-i** and in the Supplementary Movie S2 (see also calculations in Fig. S5 in the SI), where the handedness of the spiral plasmonic phase distribution is observed to switch from clockwise to counter-clockwise within each optical cycle when varying $\Delta t$. The observed periodic modulation of the helicity of the plasmonic pattern is shown in Fig. 4**j**, and shows the ability to control with attosecond precision the sign of the topological charge of the electron vortex beam created by electron-plasmon interaction.

With respect to commonly adopted methods using passive phase masks to produce electron vortex beams, our approach is intrinsically scalable to smaller length scales in the subwavelength regime (see Fig. S7 in SI). This is extremely interesting in situations where only a partial transverse coherence of the beam can be achieved (such as in the case of ultrashort electron pulses). This is even more crucial when our laser-based method for transferring OAM is extended to pulsed beams of heavier elementary and composite charged particles, where the constraint on the lateral coherence becomes stricter and limited to less than few tens of nanometers. In particular, OAM impartment in composite particles could be of interest for analyzing their internal structure (39).

When considering a proton, by shaping its wave function in a manner similar to that demonstrated here for electrons, the proton can acquire OAM-dependent density functions that will reflect its internal properties on the transverse spatial distribution. One would expect the OAM-carrying proton to behave differently in experiments whose outcome can conventionally be interpreted based only on its fixed agglomeration of partons (quark and gluons). Recently, increasing efforts have been devoted to gaining an understanding of the proton spin in terms of its constituent particles (28). Therefore, the perspective of adding OAM to the proton partons, in conjunction with relativistic spin-orbit effects in matter waves, could be of potential use to these studies (11,40). One such study we can now envisage would involve measurements of the magnetic dipole moment for an OAM-carrying proton. This quantity could be approximated based on a semi-classical current density obtained from the proton internal charge density and from the probability current density of its OAM-carrying wave function. As quantitatively described in Fig. S9 in the SI, the more the transverse component of the proton wave function approaches its internal charge density, the more the proton internal structure affects its magnetic moment. In the opposite limit, when the transverse extension of the wave function significantly exceeds the scale of the internal features of the proton, the influence of the internal charge density quickly averages out, thereby yielding a magnetic moment equal to $\ell\mu_N$ typical of a point-like particle with nuclear magneton $\mu_N$. Several schemes could be used to measure such a quantity, including a Stern-Gerlach-like approach (41) or devices that couple the OAM of a charged particle with its spin (42,43). The latter would be extremely valuable for probing the proton internal spin dynamics by means of OAM impartment, thus potentially unraveling the role played by the OAM of its inner constituents in the determination of its total spin.

**Methods**

The sample sustaining surface-plasmon polaritons (SPPs) was made of a 43 nm-thick (±5 nm) Ag thin film sputtered on a 30 nm Si3N4 membrane. A nanofabricated hole (diameter ~ 0.8 µm) was milled in the Ag layer using a raster-scanned focused-ion beam (FEI Nova Nanolab 200 focused ion beam/scanning electron microscope) with typical beam currents of about 10 pA at 30 kV voltage. The bilayer was supported by a Si TEM window with a $80 \times 80$ µm2 aperture, which was in turn mounted onto a double-tilt sample holder that ensured rotation around the x-axis by an angle α over a ±35∘ range (see also Fig. 1a in the main text).

The experiments are performed using an ultrafast transmission electron microscope as sketched in Fig. S1 in the SI and also detailed in Ref. (26) and (27). Femtosecond electron pulses are generated by photoemission from a UV-irradiated $LaB_6$ cathode, accelerated to an energy $E_0 = 200$ keV along the z axis and then directed onto the specimen surface with parallel illumination. The sample is simultaneously illuminated with femtosecond laser pulses of $\hbar\omega = 1.57$ eV central energy and variable duration, intensity, and polarization state. For the experiments presented in the main text we used a peak amplitude of light field of about 8 x $10^7$ V/m. The light pulses are focused onto the sample surface in a spot size of ~ 58 µm (FWHM). The direction of propagation of light lies within the y-z plane and forms an angle δ ~ 4.5° with the z-axis, as shown in Fig. 1a of the main text. The delay between electrons and photons is varied via a computer-controlled delay line. For the three-pulse experiment, we implemented a Michelson-like interferometer along the optical path of the near-infrared beam, incorporating a computer-controlled variable delay stage along one arm.

The transmission electron microscope was also equipped with a Gatan Imaging Filter camera for energy-filtered real-space and reciprocal-space imaging and spectroscopy. Energy-filtered real-space images presented in this work are acquired by monitoring the depletion of the zero-loss peak, following the inelastic scattering of electrons from the optically-excited plasmonic near fields (see also further details in Ref. (44)). High-dispersion diffraction (HDD) experiments have been performed at a camera length of 80 m on a retractable CCD camera before the electron energy spectrometer. In the reciprocal-space maps shown in Fig. 3 the transverse momentum is defined as 2π/distance, and the scale bar has been calibrated by static electron diffraction through a square silicon grating with lateral period of 463 nm. Both real-space images (Fig. 1, 2 and 4) and far-field images (Fig. 3) presented in the main text have been acquired in temporal coincidence between the maxima of electrons and light wave packets. During the HDD experiments we have used a micrometer-sized circular aperture (projected radius of 7.5 µm) placed in close proximity to the image plane and centered around the hole. This procedure, together with the low electron transmissivity (~ 0.013) through the continuous metal film surrounding the hole, guarantees that a significant portion (~ 30%) of the electrons reaching the detector at small diffraction angles has indeed interacted with the chiral field in the hole (see also section S4 in SI for further details).

**Acknowledgments:** The LUMES laboratory acknowledges support from the NCCR MUST. G.M.V. is partially supported by the EPFL-Fellows-MSCA international fellowship (grant agreement n. 665667). R.J.L. and D.M. gratefully acknowledge the funding support of R.J.L. by an EPSRC DTG studentship. I.K. is supported by the Azrieli Foundation, and was partially supported by the FP7-Marie Curie IOF under grant no. 328853-MC-BSiCS. V.G. is supported by the Q-SORT European project under grant agreement No. 766970. H.L. and E.K. acknowledge the Canada Research Chair (CRC) and Early Researcher Award (ERA). F.J.G.d.A. acknowledges support from the ERC (Advanced Grant 789104-eNANO), the Spanish MINECO (MAT2017-88492-R and SEV2015-0522), and the Catalan CERCA and Fundació Privada Cellex. B.B. acknowledges support on this material by the NSF under Grant No. (1759847).

**Author contribution:** G.M.V., together with F.C., F.J.G.d.A. and E.K., conceived and designed the research; G.M.V., G.B., I.M., and E.P. conducted experiments and analyzed data; R.J.L. and D.M. fabricated samples; F.J.G.d.A. developed the semi-analytical theory and performed calculations; P.B. performed the FDTD simulations; V.G. performed the electron vortex beam calculations; H.L. and E.K. performed the proton vortex beam calculations; G.M.V., G.B., I.M., O.R., I.K., B.B., V.G., E.K., F.J.G.d.A. and F.C. interpreted the results; all authors have contributed to writing the article, and read and approved the final manuscript.

**Data availability:** All data files are available from the corresponding author upon reasonable request.

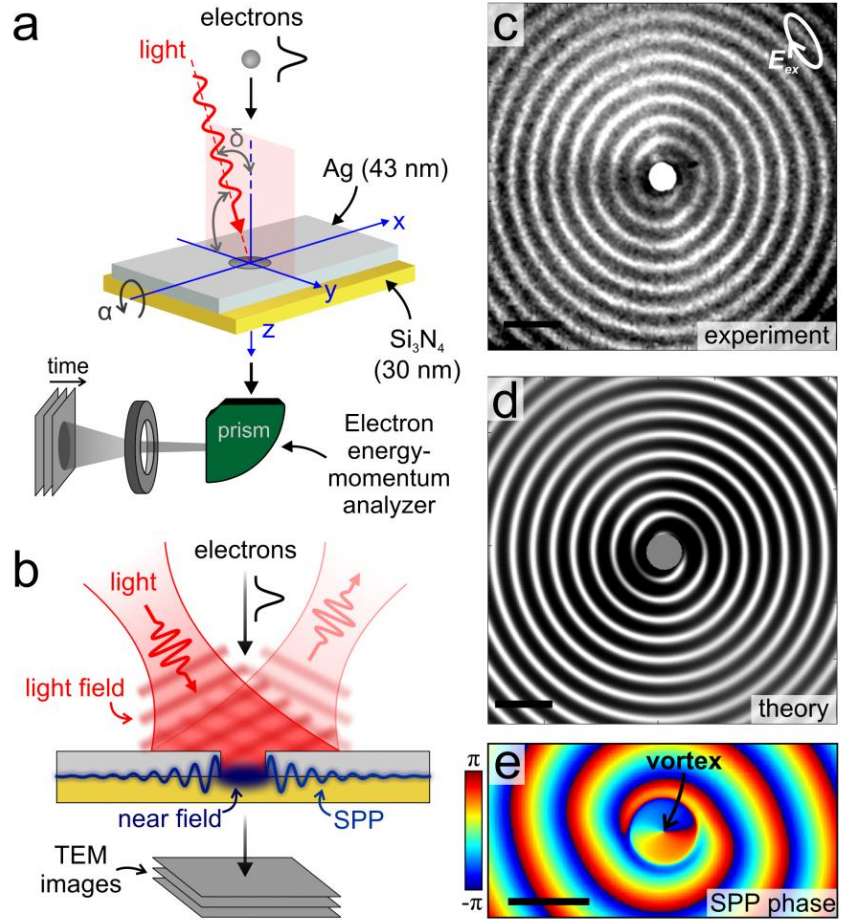

**Fig. 1**. **Generation and detection of chiral surface-plasmon polaritons.** **a** A schematic representation of the experimental geometry. Ultrashort 200-keV electron pulses (propagating along the z-axis) impinge on a Ag/Si$_3$N$_4$ thin film perforated by a nanofabricated hole in the Ag layer (diameter ~ 1 μm). Light pulses propagate with their wavevector in the $yz$ plane forming an angle $\delta = 4.5°$ with the electron beam direction. The sample can rotate around the *x*-axis by an angle $\alpha$. An electron energy-momentum analyzer allows us to measure the transverse profile of the electron wave function in both the real and the reciprocal space. **b** A schematic representation of the nonlocal holographic method used to image surface-plasmon polaritons (SPPs) radially propagating at the Ag/Si$_3$N$_4$ interface away from the hole. **c** Experimentally measured energy-filtered real-space map of inelastically-scattered electrons as a result of the electron-plasmon interaction. The image is acquired with electrons and light wave packets in temporal coincidence. The sample is tilted by an angle $\alpha = \delta$ to achieve normal light incidence. The image reveals the spiral phase pattern typical of a chiral plasmon generated by illumination with an elliptically polarized light field (the scale bar is 2 μm). **d** Calculated real-space electron intensity distribution using the semi-analytical theory developed in the text (scale bar is 2 μm). **e** Simulated phase map of the *z* component of the total electric field at the Ag/Si$_3$N$_4$ interface obtained from finite-difference-time-domain (FDTD) simulations (the scale bar is 1 μm).

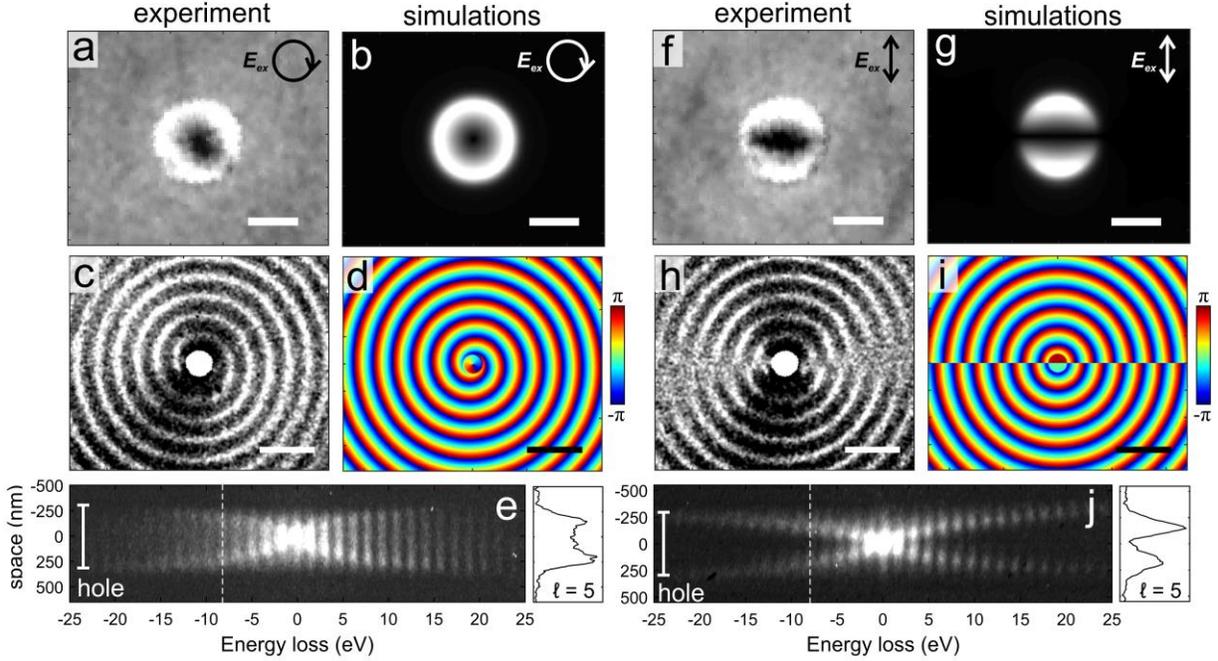

**Fig. 2. One-to-one correspondence between the complex SPP optical field and the transverse distribution of the electron wave function. a,f** Energy-filtered images of the inelastically scattered electron distribution inside the hole for circularly-polarized (**a**) and linearly-polarized (**f**) optical illumination. **b,g** Simulated spatial distributions of the interaction strength $|\beta|$ for the two polarization states. A zero-field region with cylindrical- or mirror-symmetry is visible inside the hole when circularly- or linearly-polarized light is used, respectively. **c,h** Spatially oscillating pattern measured outside the hole for the two polarizations. **d,i** Simulated spatial distribution of the phase of $\beta$ within the $xy$ plane. **e,j** Space-energy maps measured for the two polarization states (i.e., (**e**) circular and (**j**) linear) showing the modulation of the spatial distribution of the scattered electrons as a function of the number $\ell$ of absorbed/emitted SPPs.

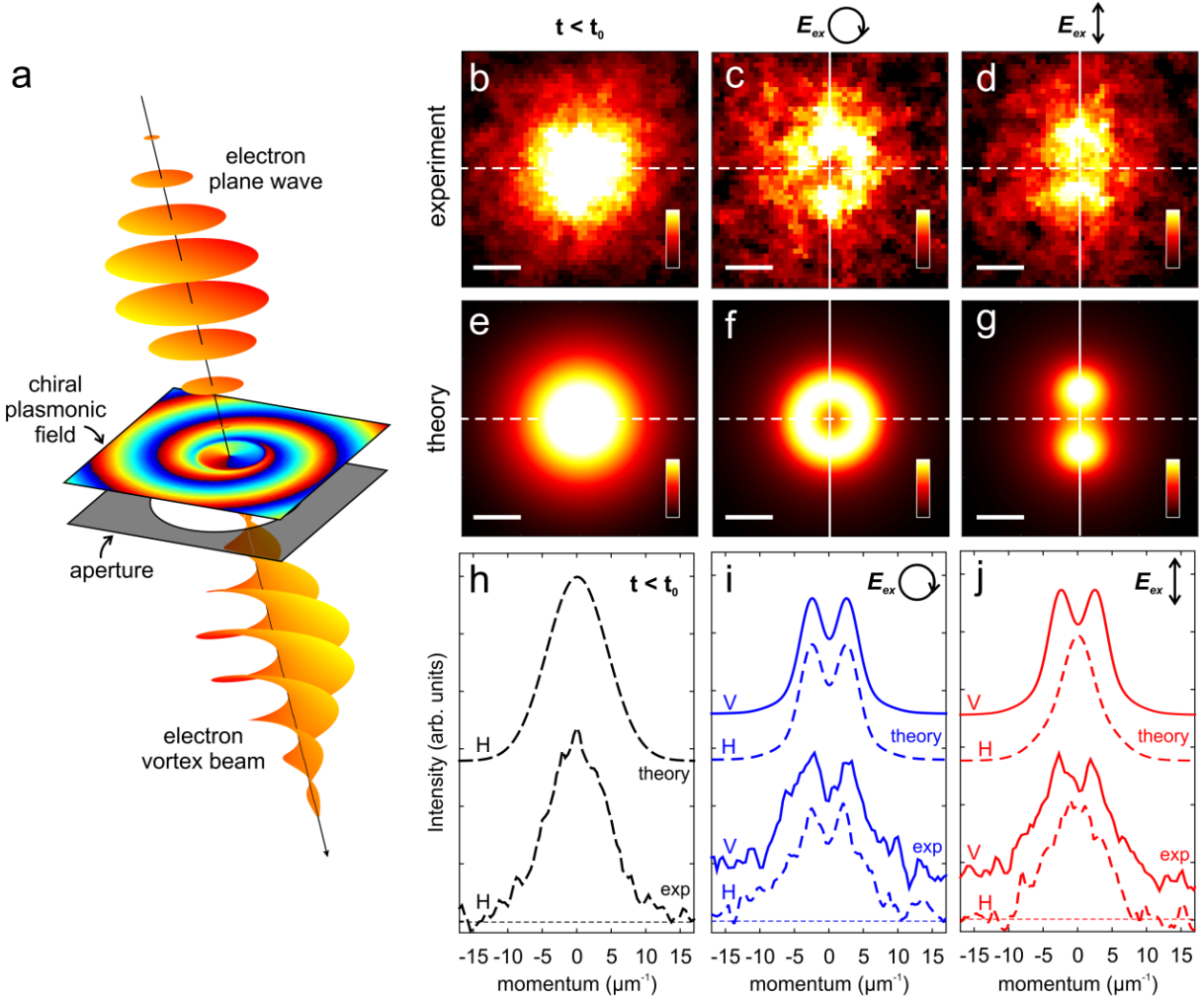

**Fig. 3. Generation and detection of an ultrafast electron vortex beam *via* interaction with a chiral plasmonic near field. a** A schematic representation of the OAM transfer to electrons: the interaction between a pulsed electron plane wave synchronized with a chiral plasmonic field creates an OAM-carrying electron wavepacket. **b**-**d** Experimentally measured far field images of the electron beam in the transversal-momentum space before the laser illumination (**b**) or in presence of circularly-polarized light (**c**) and linearly-polarized light (**d**) (the scale bar is 4 μm$^{-1}$). The images are acquired with electrons and light wave packets in temporal coincidence. Here, the experimental geometry (and in particular the angle $\alpha$) is chosen such that the interaction of the electrons with the incident and reflected light beams vanishes while maintaining a significant coupling with the plasmonic near field. **e**-**g** Calculated reciprocal space intensity distribution of the electron beam using the analytical theory and the FDTD simulations described in the text for both circular (**f**) and linear (**g**) light polarization states. The momentum-resolved electron wave function $\Psi_\ell(k_x, k_y)$ is obtained as the Fourier transform of the real-space wave function $\psi_\ell(R, \phi)$ given by Eq. (1), where the interaction strength $\beta$ originated at the hole is extracted from the FDTD simulations (the scale bar is 4 μm$^{-1}$). In panel **e**, $\beta = 0$ is considered to mimic the absence of illumination. In panel **f** and **g**, the total intensity map is shown as obtained by incoherently adding all the contributions with different values of $\ell$'s. A small momentum broadening is also included

to take into account the experimental resolution. **h-j** Vertical and horizontal line profiles across the central part of the far-field images shown in panels **b-d** (experiment) and **e-g** (theory).

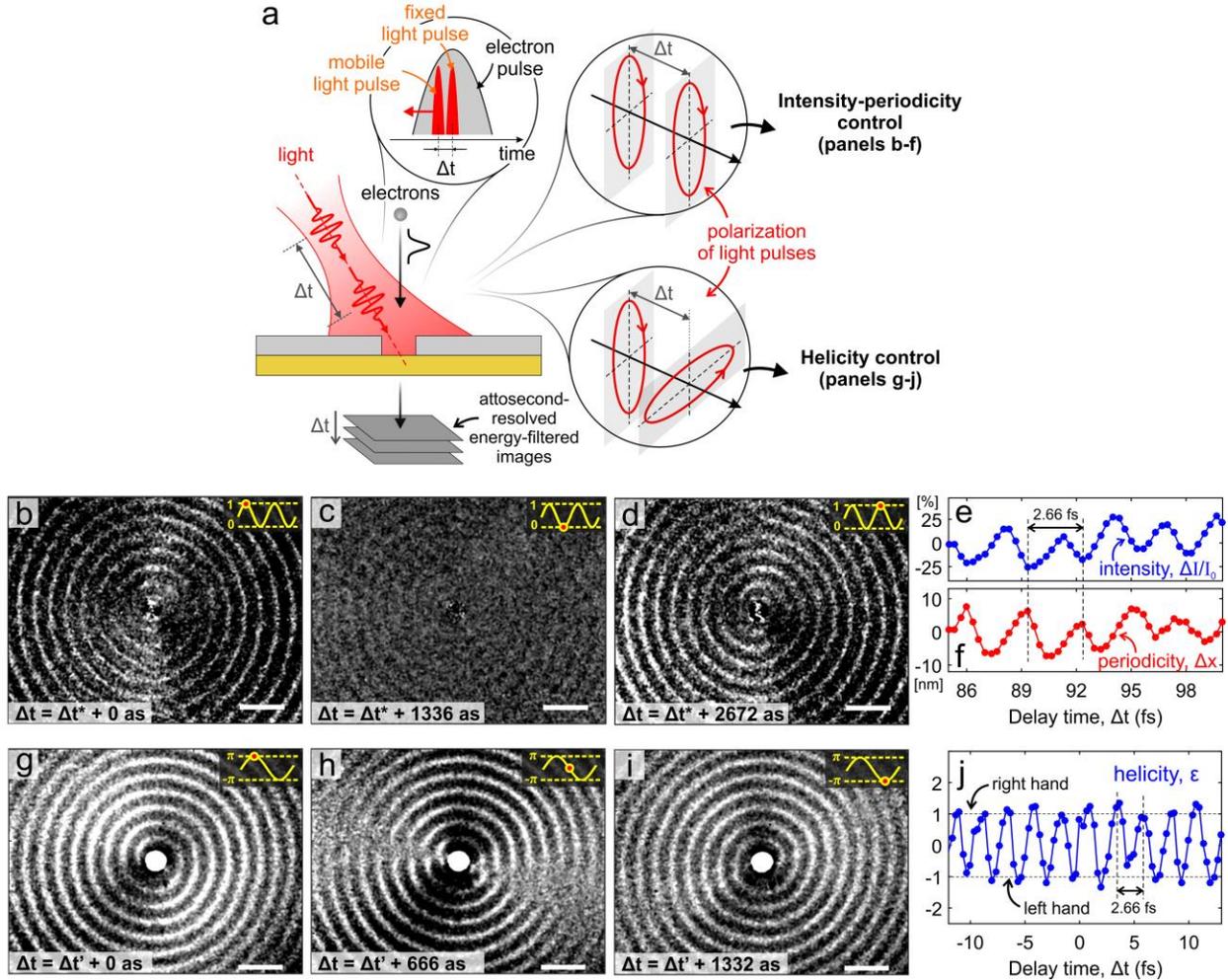

**Fig. 4. Dynamic phase control of the electron/chiral-plasmon interactions. a** A sketch of the two phase-locked light pulses with independent polarization states and variable relative delay time $\Delta t$ used to coherently launch two interfering chiral SPPs at the edge of the hole. Here, the sample is tilted to have normal light incidence ($\alpha = \delta$). **b-f** Experimentally measured energy-filtered real-space maps (panels **b-d**) of inelastically-scattered electrons as a function of $\Delta t$ (varied in steps of 334 as) when the two light pulses are elliptically polarized with parallel major axes and initially separated by $\Delta t = 85$ fs (the scale bar is 2 μm). The plots show the difference between the maps acquired at $\Delta t^* = 94$ fs and at $\Delta t > \Delta t^*$. The result is a coherent modulation of the intensity (panel **e**) and of the spatial periodicity (panel **f**) of the plasmonic fringes with a temporal period of ~ 2.67 fs. **g-j** Experimentally measured energy-filtered real-space maps (panels **g-i**) of inelastically-scattered electrons as a function of $\Delta t$ when using two elliptically polarized light pulses with major axes perpendicular to each other and initially separated by $\Delta t = -12$ fs (the scale bar is 2 μm).

The plots show the real-space maps acquired at $\Delta t' = -4.35$ fs and at $\Delta t > \Delta t'$. In this configuration, the sign of the topological charge (helicity) of the phase singularity created at the hole is observed to switch from positive to negative within each optical cycle.

**Supplementary Materials:**

Electron energy loss spectrum for electron/chiral-plasmon interaction.

Analytical theory of electron-plasmon interaction.

Finite-difference time-domain simulations and numerical determination of $\beta$.

Electron vortex beam calculations.

Parton density function of an OAM-carrying proton.

Figures S1-S9.

Movies S1-S2.